\documentclass{aa}

\usepackage[varg]{txfonts}

\usepackage{amsmath}
\usepackage{float}

\DeclareRobustCommand{\VAN}[3]{#2}
\let\VANthebibliography\thebibliography
\def\thebibliography{\DeclareRobustCommand{\VAN}[3]{##3}\VANthebibliography}

\usepackage{graphicx}	% Including figure files
\usepackage{amsmath}	% Advanced maths commands
\newcommand{\kms}{\ensuremath{\mathrm{km\,s^{-1}}}}
\newcommand{\teff}{\ensuremath{T_{\mathrm{eff}}}}
\newcommand{\msun}{\ensuremath{\mathrm{M}_{\odot}}}
\usepackage{orcidlink}

\begin{document}

\title{Violent mergers can explain the inflated state of some of the fastest stars in the Galaxy}
%\title{Inflated hypervelocity white dwarfs predominantly come from violent mergers and not the D6 scenario}

\author{Aakash Bhat\inst{1}\thanks{aakashbhat7@gmail.com}\orcidlink{0000-0002-4803-5902}
        \and
        R\"udiger Pakmor\inst{2}\orcidlink{0000-0003-3308-2420}
        \and
        Ken J. Shen\inst{3}\orcidlink{0000-0002-9632-6106}
        \and
        Evan B. Bauer\inst{4,5}\orcidlink{0000-0002-4791-6724}
        \and
        Abinaya Swaruba Rajamuthukumar\inst{2}\orcidlink{0000-0002-1872-0124}
        }
\institute{Institut für Physik und Astronomie, Universität Potsdam, Haus 28,            Karl-Liebknecht-Str. 24/25, 14476 Potsdam-Golm, Germany\\
              \email{aakashbhat7@gmail.com}
        %\and
        %    Dr Karl Remeis-Observatory \& ECAP, Friedrich-Alexander University Erlangen-Nürnberg, Sternwartstr. 7, 96049 Bamberg, Germany
        \and
            Max Planck Institut für Astrophysik, Karl-Schwarzschild-Straße 1, 85748 Garching bei München, Germany
            \and
            Department of Astronomy and Theoretical Astrophysics Center, University of California, Berkeley, CA 94720-3411, USA
        \and
            Lawrence Livermore National Laboratory, Livermore, California 94550, USA
        \and
            Center for Astrophysics | Harvard \& Smithsonian, 60 Garden Street, Cambridge, MA 02138, USA
        }

\abstract{

A significant number of hypervelocity stars with velocities between $1500-2500\,$km\,s$^{-1}$ have recently been observed. The only plausible explanation so far is that they have been produced through thermonuclear supernovae in white dwarf binaries. Since these stars are thought to be surviving donors of Type Ia supernovae, a surprising finding was that these stars are inflated, with radii an order of magnitude more than expected for Roche-lobe filling donors. Recent attempts at explaining them have combined 3-dimensional hydrodynamical supernova explosion simulations with 1-dimensional stellar modelling to explain the impact of supernova shocks on runaway white dwarfs. However, only the hottest and most compact of those runaway stars can so far marginally be reproduced by detailed models of runaways from supernova explosions. In this and a companion paper, we introduce a new \textsc{Arepo} simulation of two massive CO white dwarfs that explode via a violent merger. There, the primary white dwarf ignites when the secondary is on its last orbit and plunging towards the primary. In the corresponding aftermath, the core of the secondary white dwarf of 0.16 \msun\, remains bound, moving at a velocity of $\sim2800$\kms. We map this object into MESA, and show that this runaway star can explain the observations of two hypervelocity stars that were dubbed D6-1 and D6-3 based on their original discovery motivated by the D6 scenario, though the violent merger scenario presented here is somewhat distinct from the D6 scenario.
}
%\end{abstract}


\maketitle

%%%%%%%%%%%%%%%%%%%%%%%%%%%%%%%%%%%%%%%%%%%%%%%%%%

%%%%%%%%%%%%%%%%% BODY OF PAPER %%%%%%%%%%%%%%%%%%

\section{Introduction}

Thermonuclear supernovae in white dwarf binaries can leave behind a hypervelocity runaway companion. The discovery of the fastest stars in our Galaxy supports this scenario \citep{2018ApJ...865...15S,badry2023,hollands2025}. To explain the speeds at which these stars move ($1000-2500$ \kms) the "dynamically-driven double degenerate double detonation" (D6) scenario has been invoked. In this scenario, a white dwarf transfers mass to a more massive white dwarf through dynamically unstable Roche-lobe overflow \citep{Guillochon,Dan2011,Dan2012,2013ApJ...770L...8P,Shen2018a}, the more massive white dwarf explodes in a Type Ia supernova and leaves the secondary white dwarf behind. In fact, the discovery of the first three stars by \citet{2018ApJ...865...15S} was considered as a smoking gun for this scenario, and provided the motivation to search for them in the first place.

Since then a total of 8 candidate objects for such a scenario are known in literature. Almost all of them are inflated, having radii on the order of 0.1 R$_\odot$ instead of $<0.01$ R$_\odot$ they would have had as WDs prior to the supernova.  Relying on an \textsc{Arepo} 3D hydrodynamical simulation for the detonation phase and a calculation using the 1D stellar evolution code MESA for the long-term evolution, \citet{Bhat1} showed that the radii of some of the stars could only be inflated for a few 1000 yr due to heating by the supernova shock from the primary detonation. This is not enough to explain their observed inflated nature since kinematic ages suggest that stars were formed $0.5 - 3$ Myr ago. Recently, \citet{glanz1} found an alternative way to make fast and hot hypervelocity stars by the merger-disruption of HeCO white dwarfs in binary systems with low mass carbon-oxygen primary white dwarfs. These donors in this scenario have slightly higher velocities than what would be expected through the D6 scenario, as they are close to being disrupted, significantly overflowing their Roche lobes and plunging toward the primary white dwarf before being ejected. The 1D evolution shows these stars are inflated for times of the order $10^4-10^6$ yr.

However, while both these efforts can reproduce properties of some of the observed stars, none of them can explain the velocity and inflated nature of the cooler HVWDs (J1235, J1637, and D6 1-3). D61-3 in particular were the first three stars to be found. D6-2 is the only one whose properties are in close agreement with a He WD donor scenario \citep{sunny2024,ken2025,wong2025}, since its velocity is the lowest of all the observed stars. These stars lie in a region (\teff $<9000$ K) that none of the long-term evolutionary tracks of the previous simulations reach. 

The explanation for the high velocities of these objects in the D6 scenario relies on the assumption that these white dwarfs were Roche-lobe filling donors in white dwarf binaries. When the accretion starts, unstable mass transfer of $^4$He, leads to a thermonuclear runaway through dynamical instabilities in the interaction between the accretion stream and the shell of the accretor. When the stream is hot and dense enough, it can detonate the helium shell on the accretor. The shock from this surface detonation propagates and meets at the opposite end of the white dwarf. This shock then traverses inside to compress and consequently also ignite carbon  \citep{2007A&A...476.1133F,Guillochon, 2010ApJ...719.1067K,2013ApJ...770L...8P,2013ApJ...774..137M,2021ApJ...919..126B,pakmor2022}, completely destroying the primary white dwarf. As the timescale on which the primary white dwarf explodes is much shorter than the orbital period of the binary system, the primary suddenly disappears and the secondary white dwarf continues to move with its orbital velocity at this time.  

However, in the absence of a sufficiently massive helium shell on the primary white dwarf, it may also be possible that the accretor detonates later when the accretion stream becomes hot and dense enough to drive a carbon detonation directly. This is known as the ``violent merger'' scenario \citep{2010Natur.463...61P}. Here the donor is significantly beyond its Roche limit, so that the plunging donor star might reach higher velocities than when it fills its Roche lobe. In this case lower mass donors could have higher velocities, and such disruption-detonations could exist for $0.4$ M$_\odot$ white dwarfs \citep{Dan2012}. Whether the white dwarfs can come closer towards merging and not detonate before that depends on the mass fraction and the $^4$He shell masses of the two white dwarfs \citep{shen2024}. If the white dwarfs could come close to merging and the donor could survive the impending supernova explosion, then this would help alleviate some tension between the ejection velocities of the white dwarfs and their masses, since lower mass white dwarfs could also achieve $\sim2000$ \kms velocities.

In this work we simulate the long-term evolution of a runaway surviving remnant from the violent merger scenario, and show that in comparison to the D6 scenario, the runaway can achieve both higher runaway velocity and a more inflated radius for Myr timescales.
This paper is structured along with a companion paper, where we report on a new 3-dimensional hydrodynamical simulation of binary white dwarfs which undergo a supernova explosion and end with a hypervelocity runaway donor remnant. We simulate this using the 3D magneto-hydrodynamic code \textsc{Arepo} \citep{2010MNRAS.401..791S,Pakmor2016,2020ApJS..248...32W}. The simulation followed the evolution until shortly after detonation of the primary white dwarf. We take this surviving donor as input into the 1-dimensional open-source stellar evolution software Modules for Experiments in Stellar Astrophysics (MESA, \citealt{Paxton2011,Paxton2013,Paxton2015,Paxton2018,Paxton2019,Jermyn2023}).We use the entropy profiles from \textsc{Arepo} to relax and model the subsequent stellar evolution through the following $100\, \mathrm{Myr}$, similar to what was done in \cite{Bhat1} and \cite{glanz1}. In Section~2, we discuss the \textsc{Arepo} simulation. In Section 3, we discuss relaxation in MESA. Section 4 discusses the long term evolution of the models and section 5 compares them to recent observation of D6 runaways. We summarize our results in Section 6. 

\section{3D Hydrodynamical Simulation}

In this section, we summarise the details of the explosion simulation that produces the bound donor remnant that we then evolve further in MESA. We refer the reader to the companion paper (Pakmor et al, in prep) for a comprehensive explanation of the simulation. We start in \textsc{Arepo} with a double white dwarf system. This simulation starts with the late stages of the inspiral of a binary system of two carbon-oxygen white dwarfs with masses of $1.10\,\mathrm{M_\odot}$ and $0.70\,\mathrm{M_\odot}$, respectively. The donor model starts with a 50/50 C/O core composition, $0.016\,\mathrm{M_\odot}$ $^{22}$Ne, and a $^4$He shell of $0.0003$\msun. The secondary has $0.32\,\mathrm{M_\odot}$ carbon, $0.36\,\mathrm{M_\odot}$ of oxygen, $0.011\,\mathrm{M_\odot}$ of $^{22}\mathrm{Ne}$, and a helium shell of $0.009\,\mathrm{M_\odot}$. While helium is accreted in the beginning of the simulation, this helium is expelled from the system in surface detonation-powered outbursts. Due to the small He shell on the accretor, no coherent helium detonation forms on the primary white dwarf, so the double detonation scenario is avoided. After a few orbits, we artificially decrease the separation to avoid computational costs associated with orbital shrinking. When the white dwarfs are close enough for the secondary to be disrupted, there is almost no $^4$He left in the system. In the last inspiral of the system, the donor is accelerated towards the accretor. Its impact with the surface of the accretor is enough to create conditions for direct carbon ignition, and the primary white dwarf explodes. The orbital velocity of the donor before it hits the accretor is $\sim2700$ \kms.

The accretor explodes completely. In contrast to previous violent merger simulations \citep{2010Natur.463...61P,Pakmor2012b,2015ApJ...807..105S}, a small amount of the donor white dwarf ($\sim 0.16$ \msun) remains bound after being hit by the explosion of the primary white dwarf and being burned partially. A small amount of $^{56}$Ni and carbon burning ashes are deposited from the low-velocity tail of the supernova ejecta. Around $1.64$ \msun\, of material is ejected in the explosion.  
\begin{figure*}
\centering
\includegraphics[width=0.98\textwidth]{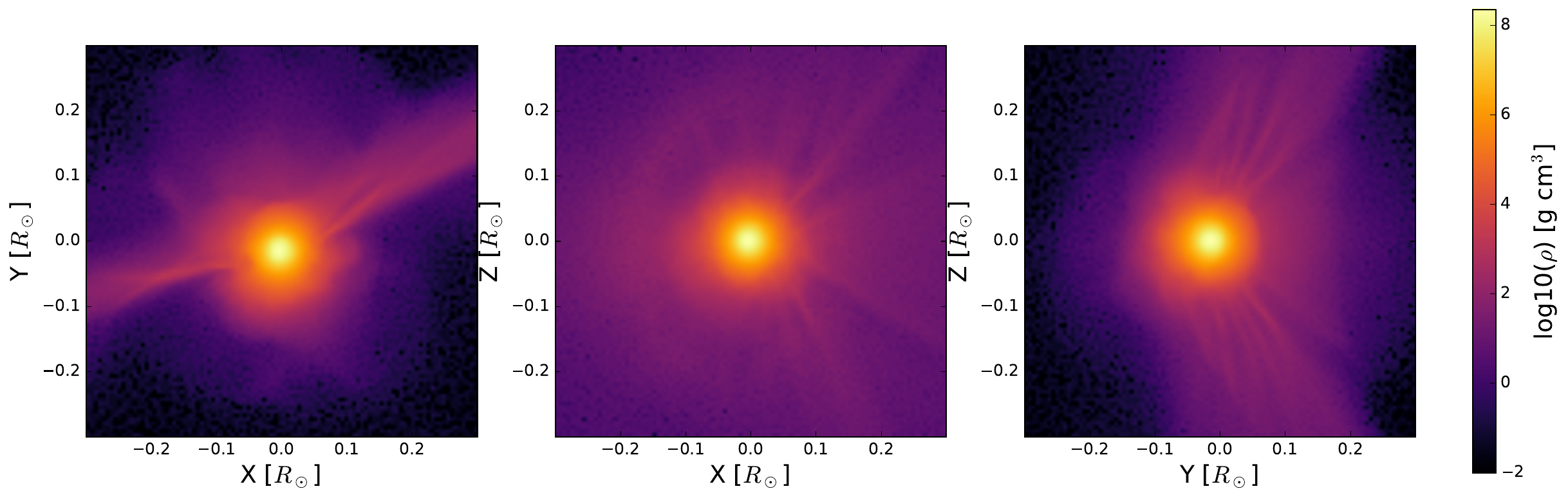}  \caption{Planar slices of the runaway star density 1000 seconds after explosion. A slight asymmetry and a non-spherical extension of the ejecta is visible outside of the inner $0.05$ R$_\odot$.}
\label{fig:3d}
\end{figure*}

We use \textsc{Arepo} to follow the expanding supernova ejecta and the bound surviving donor star until $1000\,\mathrm{s}$ after the explosion. This is of the order of a few dynamical timescales for our slightly expanded runaway,
\begin{equation}
\tau_\mathrm{dyn} = \sqrt{\frac{R^3}{G M}} 
\sim \sqrt{\frac{(0.1\,R_\odot)^3}{G \,(0.16\,M_\odot)}} 
\approx 120\,\mathrm{s}
\end{equation}

\begin{figure}
\hspace*{-0.3cm}
\centering
\includegraphics[width=0.5\textwidth]{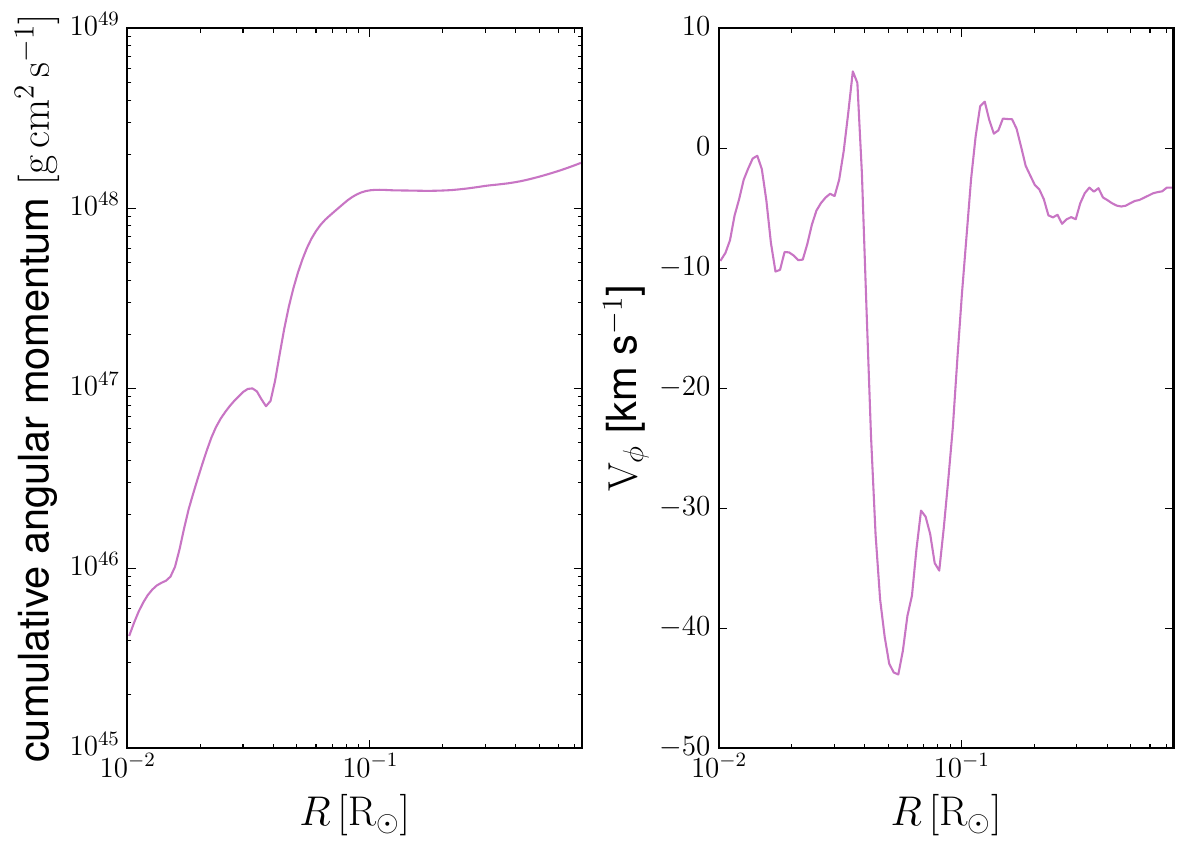}  
\caption{The total angular momentum and the rotational velocity of the runaway at around $1000 \,s$ after the explosion. The plotted quantities are spherical averages and highlight the differences in the state of the stellar interior. Most of the angular momentum is within $0.1$ $R_\odot$. While the external ($>0.1$ $R_\odot$) has some residual rotation and radial energy this is only in a small region which forms a stream like structure.}
\label{fig:rem}
\end{figure}

\begin{figure}
\hspace*{-0.3cm}
\centering
\includegraphics[width=0.5\textwidth]{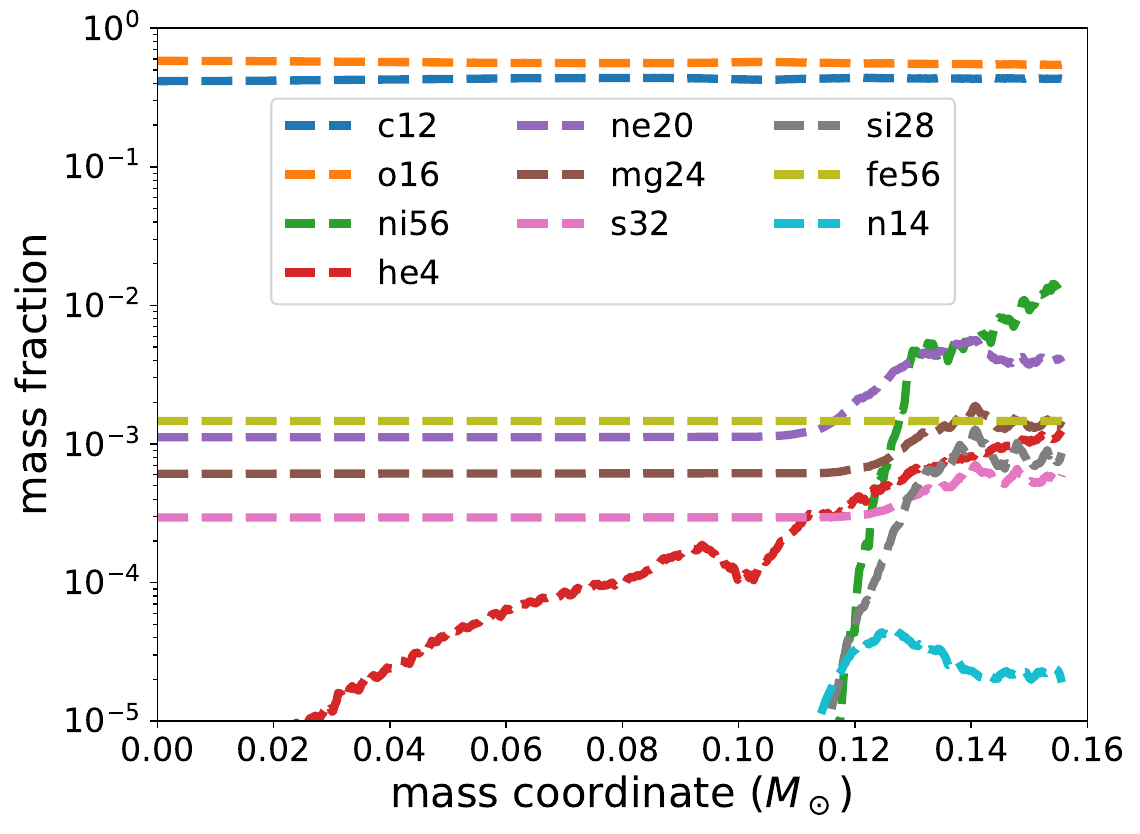}  \caption{Main elements as a function of mass coordinate for the runaway 1000~s after the supernova for mass coordinates which remain bound after supernova explosion. Unlike in \citet{Bhat1}  $^{56}$Ni does not dominates the mass fraction at the surface of the runaway. $^{56}$Fe is present due to the primordial solar abundance assumed.  }
\label{fig:compo}
\end{figure}

The final \textsc{Arepo} profile of the remnant that is mapped into MESA contains composition information and averaged 1-D temperature-density pairs for different shells of the star. 2D slices of the 3D density structure of the runaway are shown in Figure \ref{fig:3d}. The total angular momentum and the corresponding averaged rotational velocity are shown in Figure \ref{fig:rem}. While the inner structure ($<0.05$$R_\odot$) does not deviate more than $10$\,\kms in rotational velocity, the outer part between $0.05-0.1$ $R_\odot$ has slightly higher rotational velocity variations, deviating from rigid body rotation. However, computational limits did not allow us to go farther than this in 3D. The final composition profiles of the main elements with mass fractions $\max(X_n) > 10^{-5}$ in the runaway are shown in Figure \ref{fig:compo}. Unlike the runaway in \citet{pakmor2022}, the overall contamination of the surface of the runaway is significantly less, with $^{56}$Ni only one-hundredth of the mass fraction at the surface. In the next sections, we use this surviving donor of this simulation as a starting point to study its subsequent evolution. We employ the composition and entropy profiles of the donor at the end of the \textsc{Arepo} simulation and evolve them further in MESA. As seen in Figure \ref{fig:3d}, the runaway star has asymmetric plumes most likely resulting from the pre-supernova configuration of the donor and its interaction with the shock. The exact modeling of this plume is not possible in 1D, and there is a possibility that this part of the star is unbound on timescales longer than what can be followed in \textsc{Arepo}. Therefore, we motivate different stellar masses in MESA by cutting out regions where the ejecta is non-spherical in \textsc{Arepo} to test our models.

\section{Mapping the hydrodynamical result into the 1D stellar evolution code MESA}

\begin{table}[h!]
\centering
\begin{tabular}{lrrrrr}
\hline
\textbf{Total (\msun)} & \textbf{$^{56}$Ni (\msun)} & \textbf{ $^{4}$He (\msun)} & \textbf{ $^{12}$C (\msun)} & \textbf{$^{16}$O (\msun)} \\
\hline
0.154 & 7.0 $\times$ 10$^{-5}$ & 2.0 $\times$ 10$^{-5}$ & 0.065 & 0.087 \\
0.140 & 1.8 $\times$ 10$^{-6}$ & 8.8 $\times$ 10$^{-6}$ & 0.059 & 0.079 \\
0.120 & 1.6 $\times$ 10$^{-10}$ & 3.3 $\times$ 10$^{-6}$ & 0.050 & 0.068 \\
0.100 & 2.4 $\times$ 10$^{-11}$ & 1.0 $\times$ 10$^{-6}$ & 0.042 & 0.057 \\
\hline
\end{tabular}
\caption{Mass fractions of key isotopes in the remnants for varying total remnant masses. The table lists the total remnant mass along with the masses of $^{56}$Ni, $^4$He, $^{12}$C, and $^{16}$O in solar masses.}
\label{tab:remnant_composition}
\end{table}

To model the long-term evolution of the remnant runaway, we evolve it in MESA \citep{Paxton2011, Paxton2013, Paxton2015, Paxton2018, Paxton2019, Jermyn2023}, an open-source 1D stellar evolution code. The components of the MESA EOS blend relevant for this work are FreeEOS \citep{Irwin2004} near the surface of the models, and HELM \citep{Timmes2000} and Skye \citep{Jermyn2021} for white dwarf cores.
MESA uses tabulated radiative opacities primarily from OPAL \citep{Iglesias1993,
Iglesias1996}, with low-temperature data from \citet{Ferguson2005}
and the high-temperature, Compton-scattering dominated regime by
\citet{Poutanen2017}. The electron conduction opacities are from
\citet{Cassisi2007}. Nuclear reactions are from JINA REACLIB \citep{Cyburt2010}, NACRE \citep{Angulo1999} and additional tabulated weak reaction rates \citet{Fuller1985, Oda1994,Langanke2000}. Screening of reaction rates is included via the prescription of \citet{Chugunov2007}, while thermal neutrino loss rates are from \citet{Itoh1996}. The inital white dwarf models are made with the make\_co\_wd test suite, with solar composition of \citet{asplund}. The white dwarf masses are then relaxed to the masses of our models using the \texttt{relax\_mass} hook.
\begin{figure}
\hspace*{-0.3cm}
\centering
\includegraphics[width=0.5\textwidth]{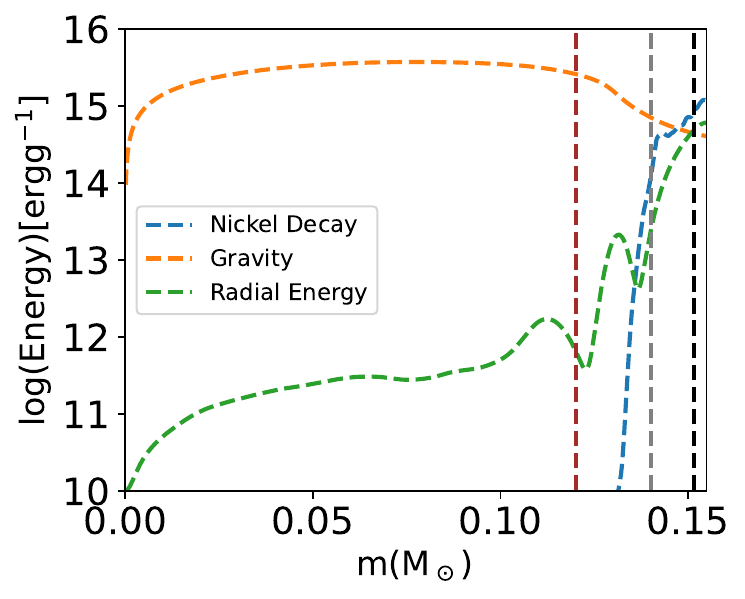}  \caption{Different energies in the runaway WD as function of its mass coordinate. Black line marks the region where nickel decay dominates, grey line marks the region where gravity dominated over other energies, and brown line marks the region where gravity is the only relevant energy. Rotational energy is not shown as it is negligible. Radial energy is defined as $v^2_r/2$. }
\label{fig:energy}
\end{figure}

In previous work by \citet{Bhat1}, the change in thermal state of the post-explosion runaway white dwarf donor compared to the pre-explosion donor was used to map the change in {\em entropy} that the donor star experiences as a result of interaction with the supernova ejecta. In this paper, the final runaway star configuration is not only due to supernova stripping and shock heating (as was the case in \citealt{Bhat1}), but is also affected by the disruption of the secondary white dwarf just before the explosion (as was the case in \citet{glanz1}). That is, we believe the thermal state of this runaway is also dominated by the hydrodynamics just before the supernova, which are captured in the Arepo model. They are less sensitive to the thermal state of the model prior to interaction with its companion, and we can therefore map the Arepo entropy more directly into MESA, rather than relying on an entropy difference. Furthermore, defining an entropy difference is more complicated for this system, since the donor loses more than $75\%$ of its mass. Since MESA is a 1-D evolutionary code, we average the 3-D profiles from \textsc{Arepo}. We employ the averaged profiles directly by repurposing the MESA relaxation test suite \texttt{relax\_composition\_j\_entropy} to initialise a WD model directly to the entropy and composition profile specified by the final state of the \textsc{Arepo} model. Here, the procedure followed is similar to that outlined in \citet{glanz1}. Since the overall mass loss is too high, and the white dwarf is almost completely disrupted, we do not use the heating approach used in \citet{Bhat1} to generalize the heating for multiple white dwarfs. For this work, we do not use any external opacity tables and rely on the Type II opacities in MESA. We neglect diffusion for now. Since these stars are cooler and have lower surface gravities compared to the new D6 stars discovered by \citet{badry2023}, we do not expect diffusion to play a big role for the time that we evolve them.

Since the boundary of the post-explosion runaway is difficult to average (due to the non-spherical nature of the bound object as shown in Figure \ref{fig:energy}), we average on equipotential surfaces of the gravitational potential. This is different than the spherical averaging employed before by \citet{Bhat1} and \citet{glanz1}, and is preferred because our system is non-spherical. Furthermore, we define four models with different cuts to the outer parts of the profile. These are defined by the approximate energy regimes as shown in Figure \ref{fig:energy}. The first cut is defined by the decay energy of $^{56}$Ni being more than twice the gravitational energy. This is $0.154$ \msun. The second cut is defined by the limit that the Arepo model is spherically symmetric such that the specific radial energy, defined as $v^2_r/2$ where $v_r$ is the averaged velocity in the radial direction, is less than $1\%$. This model has a mass of $0.12$ \msun. The third cut is chosen in between the first two masses around $0.14$ \msun\, where the gravitational energy dominates but the kinetic and decay energies are still significant. The final cut is a lower limit at $0.10$ \msun. This model has no significant ($>10^{-5}$ in mass fraction) supernova pollution. These mass cuts allow us to evolve multiple stars with different amounts of elements at the surface. Since we skip the dynamical evolution of the remnant before it reaches full spherical hydrostatic equilibrium due to computational limitations, spherical averaging is only approximate in the outer regions of the remnant. We can therefore capture some uncertainty that is associated with the 3D averaging by making these mass cuts.

From the theory of tides, one would expect the donor white dwarf to rotate extremely fast (on the order of $\sim 40$ seconds) to come close to the orbital period \citep{Marsh2004,Fuller2012,fuller2014,yu2020}. However, due to the tidal distortion pre-supernova and the total disruption by the supernova with the corresponding expansion of the surviving star, the post-supernova runaway shows almost negligible rotation. As the star would contract over time this rotation rate should increase but will not be significant. Much of the angular momentum of the system is lost in the disruption and stripping all but the very core of the donor white dwarf. We therefore do not consider any rotation in our models. Apart from D6-2, which rotates with a period of 15.4 hours \citep{2022MNRAS.512.6122C} but most likely does not come from this scenario, the other runaways are not yet known to harbour significant rotation. J1637 has been confirmed to have an upper limit of $40$ \kms\, by \citet{hollands2025}, based on the spectral resolution and stellar line broadening.

For all evolution and relaxation scenarios, we use the {\tt approx21\_plus\_co56} nuclear network. This includes the isotopes $^{1}$H, $^{3,4}$He, $^{12}$C, $^{14}$N, $^{16}$O, $^{20}$Ne, $^{24}$Mg, $^{28}$Si, $^{32}$S, $^{36}$Ar, $^{40}$Ca, $^{44}$Ti, $^{48,56}$Cr, $^{52,54,56}$Fe, $^{56}$Co, and $^{56}$Ni.

\section{Long term evolution}
\label{s.Evolution}

As described in \citet{Bhat1} and \citet{glanz1} we carried out the subsequent evolution after relaxation to a maximum age of 100 Myr. We enable nuclear burning, convection, and thermohaline mixing \citep{1980A&A....91..175K}. Compared to the WDs produced by the D6 scenario in previous works \citep{pakmor2022}, there is much less radioactive material here. Nevertheless, the atmosphere is still slightly polluted. We also switched on super-Eddington winds for mass loss driven by any $^{56}$Ni decay, but found that at no time is this energy enough to unbind any surface layers. 

\begin{figure}
\centering
\includegraphics[width=0.5\textwidth]{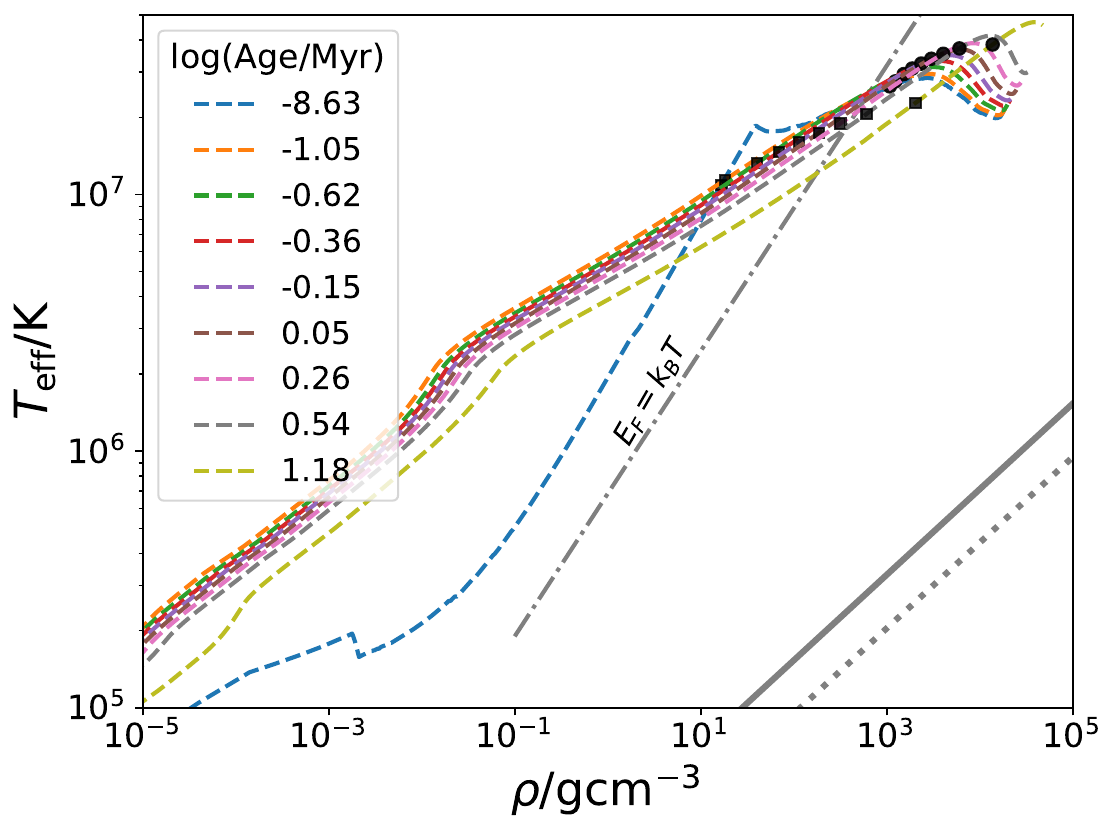} % Change the filename and options accordingly
\caption{The structural profile of the $0.154M_\odot$ model as a function of its age. The puffing up and the later contraction of the star is seen here. The black circles (black squares) represent $50\%$ ($90\%$) of the stellar mass. The grey dashed-dotted line is the degeneracy limit above which gas pressure dominates. The two grey lines in the bottom right are the crystallization limit for carbon and oxygen. }
\label{fig:rho_T}
\end{figure}

The evolution of the 0.154 \msun\, model is slightly similar to what was seen for the D6 runaways in \citet{Bhat1}, but in a completely different phase space of the HR-diagram. In particular, our models are close to the boundary of degeneracy, even in the stellar core. For this reason, we do not consider the runaway stars to be white dwarfs at this particular stage of their evolution. The evolution of the $\rho-T$ profile is shown in Figure \ref{fig:rho_T} for this model. As the star was almost completely disrupted, the density of the star is significantly lower than that of the pre-supernova donor. As such, for the early stages of the evolution, almost $50\%$ of the stellar mass is close to the degeneracy limit given by the Fermi energy being equal to the thermal energy, $E_F = k_B T$. The outer $10\%$ of the star by mass is non-degenerate. Over the first $10^5$ yr this region of the star expands while the inner region contracts and becomes hotter. The entirety of the star becomes close to completely degenerate by the time $\sim10$ Myr have passed. Our models only come close to their corresponding cooling track at the end of our simulations (at 100 Myr). This rho-T space shows the evolution to a configuration similar to the lowest mass models in \citet{ken2025} for the external $10\%$ of the star. The total initial configuration is however different since the models of \citet{ken2025} are purely convective, have no degenerate material, and are significantly cooler.

While this model does not have significant $^{56}$Ni, some decay still occurs to produce iron. For $7\times10^{-5}$ \msun\, of $^{56}$Ni this amounts to around $10^{46}$ erg.  Within our models, the highest specific luminosity is of the order $\sim 10^8$ erg/g/s at the beginning of the evolution. This is enough to heat the runaway star significantly to expand, but, as this drops exponentially, the heating is not sustained long enough to lose any mass due to the Eddington luminosity limit. For our models, this limit is close to $L_\mathrm{Edd} = \frac{4 \pi G M c}{\kappa} \sim 10^4 L_\odot$. Such a luminosity is only produced in the first few days of the nickel decay. There is a corresponding expansion of the star as it thermally re-adjusts. 

Since the MESA thermohaline mixing procedure does not capture the impact of the heavier metals in the surface layers where the FreeEOS tables are called, we also apply a minimal amount of mixing in the outer $\sim3\%$ of the mass of the star, similar to what was done in \citet{Bhat1}.

For all the other models where no significant amount of supernova contamination exists due to surface cuts, the evolution of the runaway star is very similar to this model post $1000$ yr (that is, after the nickel has decayed). For those models the nuclear decay energy is only of the order $10^2-10^4$ erg/g/s for the first $\sim100$ days and decays exponentially. The evolution of these models is dominated by thermal adjustment due to tidal heating near disruption and the heat injected by the supernova shock, until the subsequent contraction and cooling towards the white dwarf cooling track.

\section{Comparison with observations}

\begin{figure}
\centering
\includegraphics[width=0.45\textwidth]{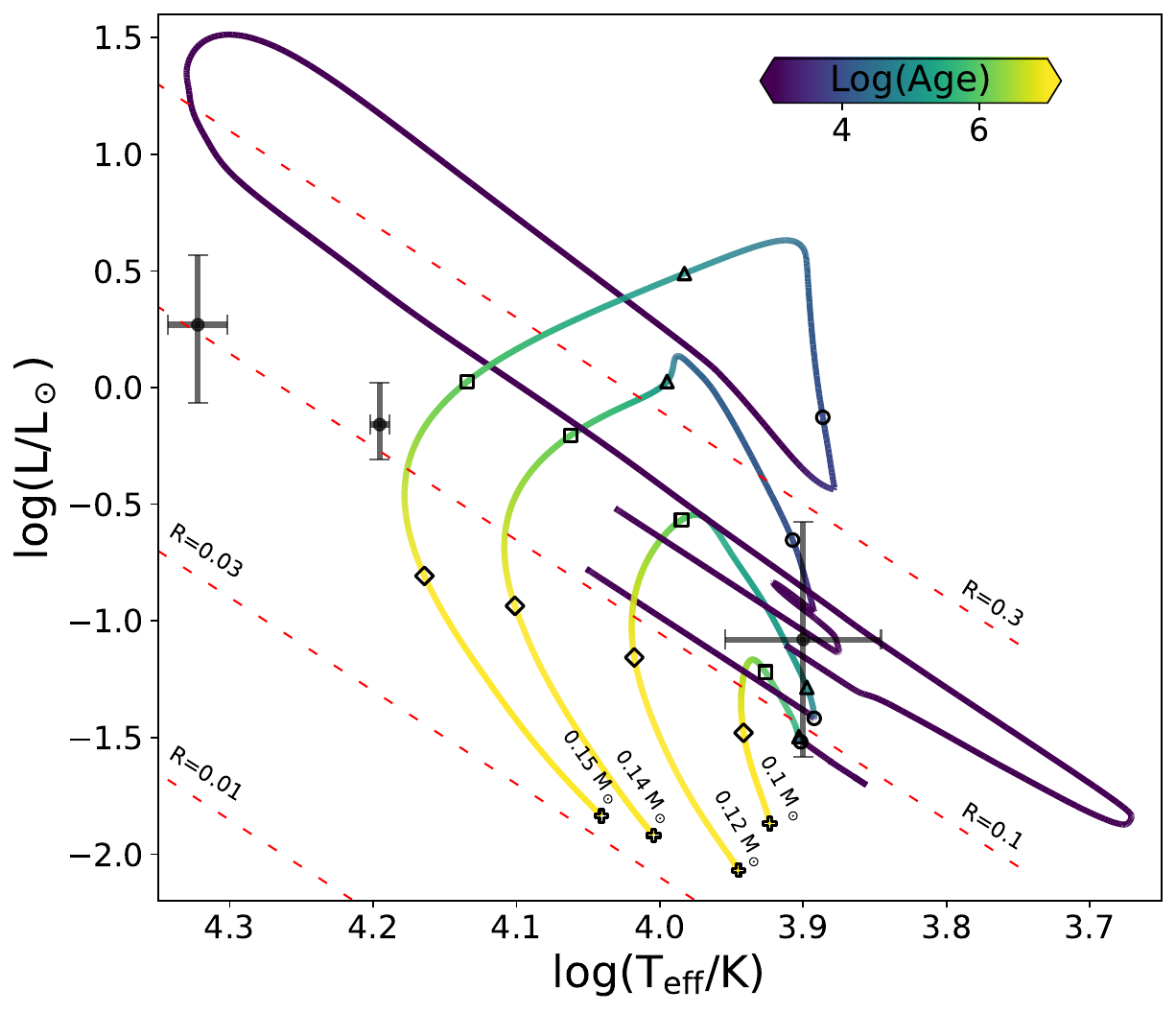}
\includegraphics[width=0.45\textwidth]{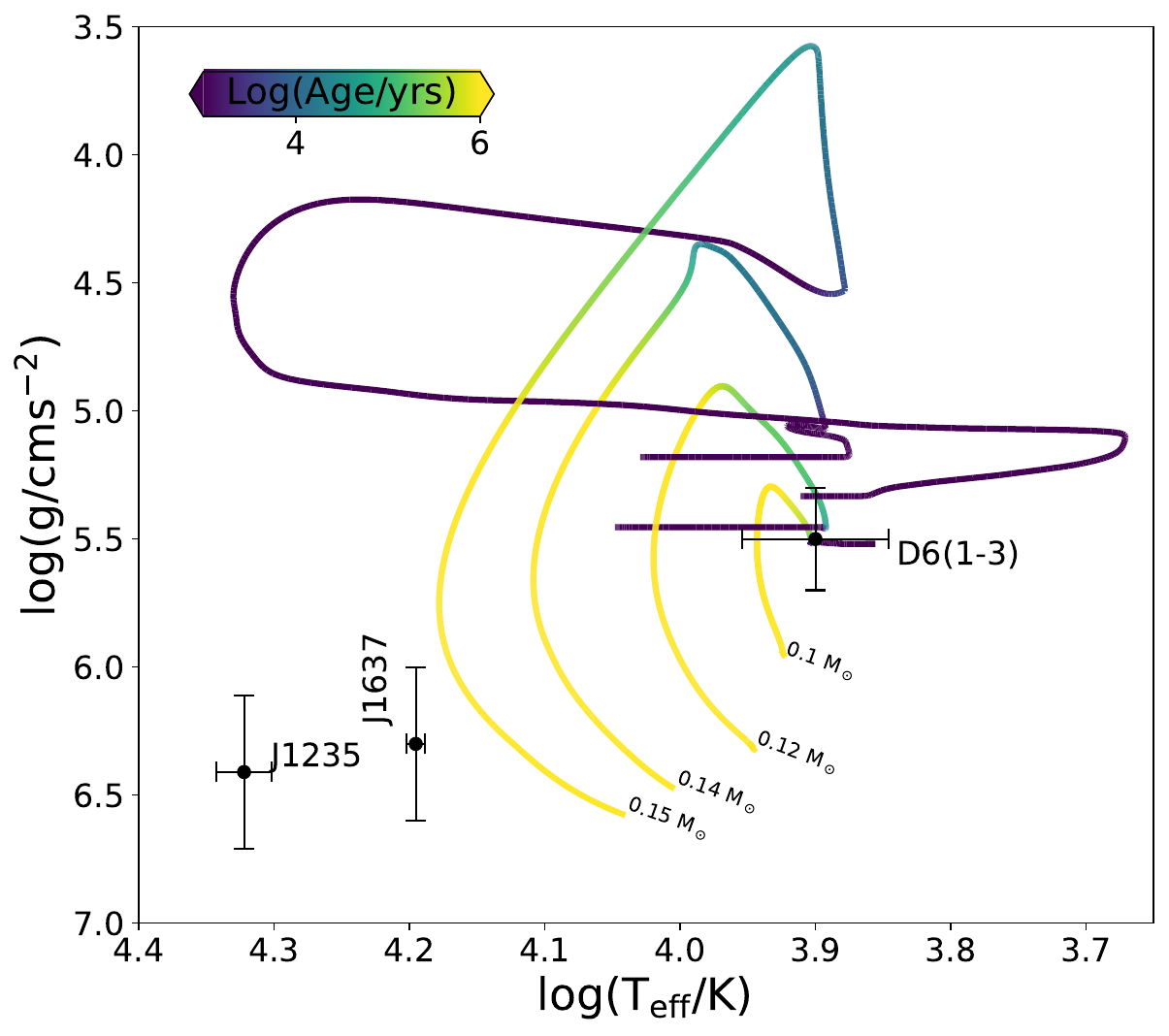}
\caption{\textbf{[upper panel]} HR diagram for all surface cuts. All observed D6 stars are plotted as black points. Dashed red lines represent lines of constant radii. Colorbar shows the log(age) of the star. Markers show the ages at $10^4, 10^5, 10^6, 10^7, 10^8$ years. \textbf{[lower panel]} The Kiel diagram for the same model. }
\label{fig:kiel}
\end{figure}

The Hertzsprung-Russell (HR) and Kiel diagrams of all the models are shown in Figure \ref{fig:kiel}, with the D6 observations over-plotted.  For a discussion on the values of radius, surface temperature, and surface gravity of all except one object (J1637) we refer the reader to \citet{Bhat1}. We omit the hottest star J0546 because it is far outside the parameter range of our models and instead plot the new D6 star J1637 \citep{hollands2025} which fits better here. For J1332 we consider the hydrogen-dominated model from \citet{werner2024}. Although the presence of this hydrogen comes with caution \citep{badry2023,glanz1}, it may be explained by Bondi-Hoyle-Lyttleton accretion of $\sim 10^{-16}$ \msun\, hydrogen from the ISM \citep{werner2024,ken2025}. The luminosity was calculated using:
\begin{equation}
(L/L_\odot) = (R/R_\odot)^2\left(\frac{T_{\text{eff}}}{5800\,\rm K}\right)^4~.
\label{eq:bb}
\end{equation}
Unlike the models of \citet{Bhat1}, the models here do not have significant radioactive nickel, due to which the stars cool in the beginning with a roughly constant radius. This process is disrupted by the internally deposited heat of the star (and in the case of one model, the nickel decay), making its way to the surface as the runaway star starts to expand and increase its luminosity. The evolutionary tracks of our lower mass models pass through the positions of D6-1 and D6-3, showing for the first time that survivors of violent merger might be able to explain these stars.

\subsection{Radius and $T_{\rm eff}$}

\begin{figure}
\centering
\hspace*{-0.2cm}
\includegraphics[width=0.45\textwidth]{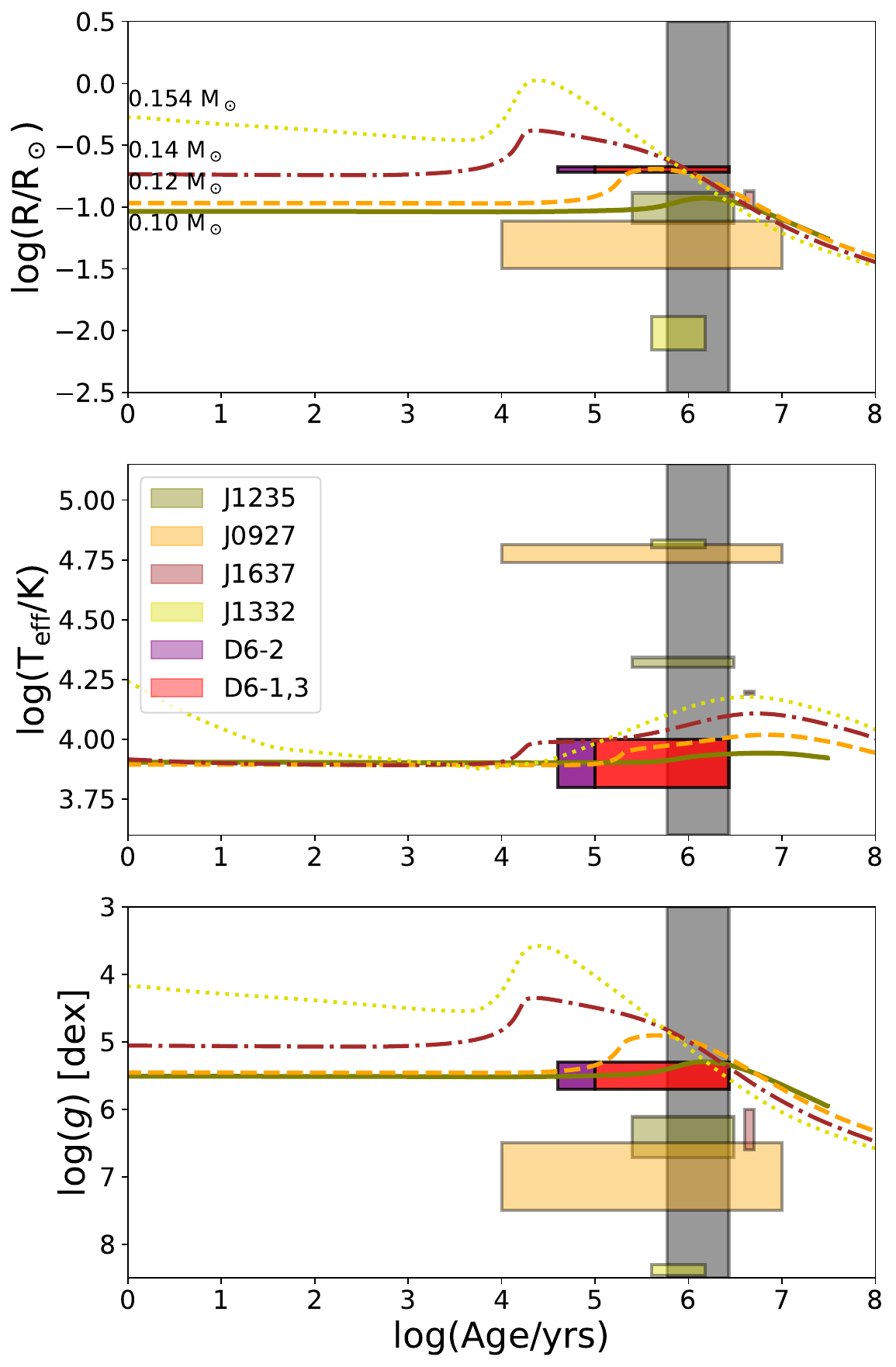}
\caption{Evolution of radius, effective temperature, and surface gravity of our models. D6 observations are plotted here for the radius. While the radii of D6-1 and D6-3 are similar to those of D6-2, their age estimate is less certain and spans a bigger region. }
\label{fig:lrt_2}
\end{figure}

A more precise comparison with the observed values is needed to gage whether the expansion occurs at a timescale similar to that of the observational timescale of these stars. The evolution of radius, effective temperature, and surface gravity of our models over time is shown in Figure \ref{fig:lrt_2}. The models puff up on the order $10^4$ yr for the highest mass model to $10^6$ yr for the lowest mass model. This is seen in the radius and surface gravity plots. The increase in surface temperature corresponds to the contraction after the star has puffed up to its maximum radius. We also over-plot the relevant D6 observations again along with their kinematic ages taken from \citet{Bhat1}.  

To understand the relation of the kinematic age and the expansion of the star, we can look at the relevant timescales. The expansion and heating of the stars is governed by two thermal timescales. The local thermal timescale is given as \begin{equation}
    t_{\rm{th}}=\frac{H^2}{D_{\rm{th}}}~,
    \label{eq:td}
\end{equation}
where $H=P/\rho g$ is the local pressure scale height and $D_{\rm{th}} = 4 a c  T^3/3 \kappa\rho^2 c_P$ is the coefficient of thermal diffusion. This timescale is relevant for the local heat transfer within any star and sets the time for the runaway star to transfer the internal heat to the surface allowing the star to readjust. In comparison, the Kelvin-Helmholtz timescale sets the timescale of cooling of the runaway star once this heat has been fully transferred outside. This is given by
\begin{equation}
    t_{\rm{kh}}=\frac{GM^2}{R L}~.
    \label{eq:td}
\end{equation}

\begin{figure}
\centering
\includegraphics[width=0.45\textwidth]{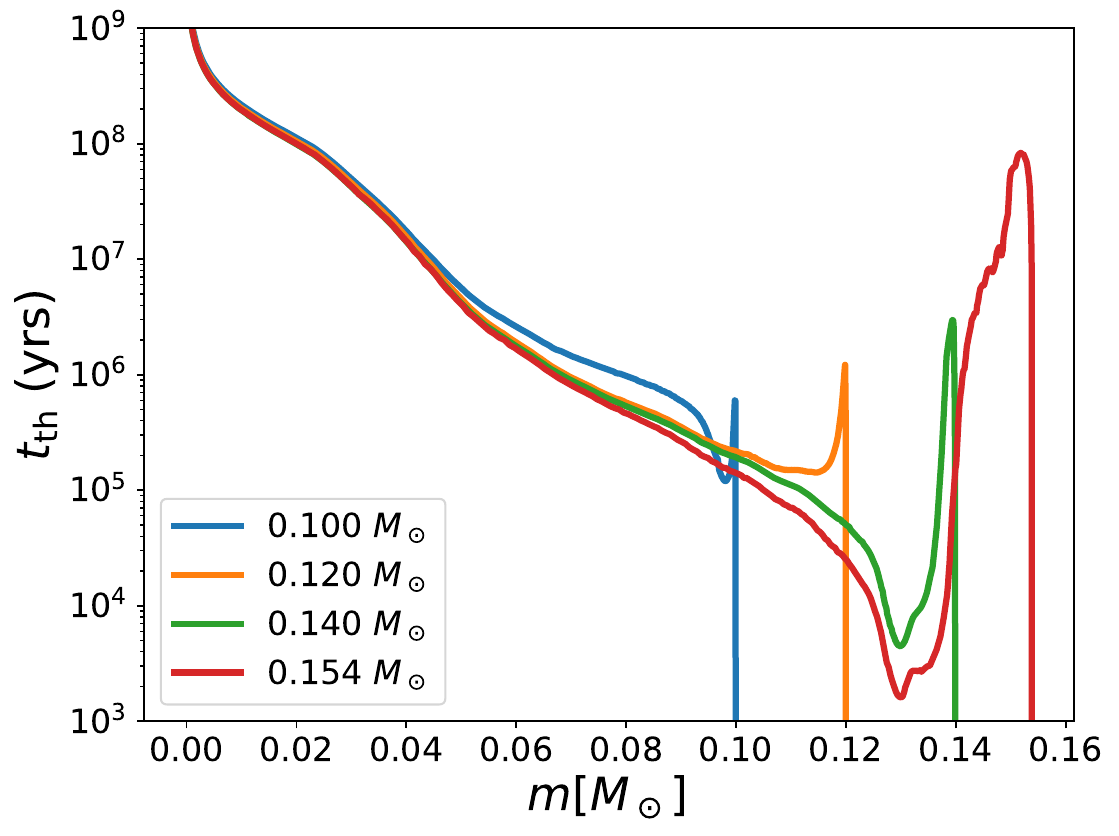}
\includegraphics[width=0.45\textwidth]{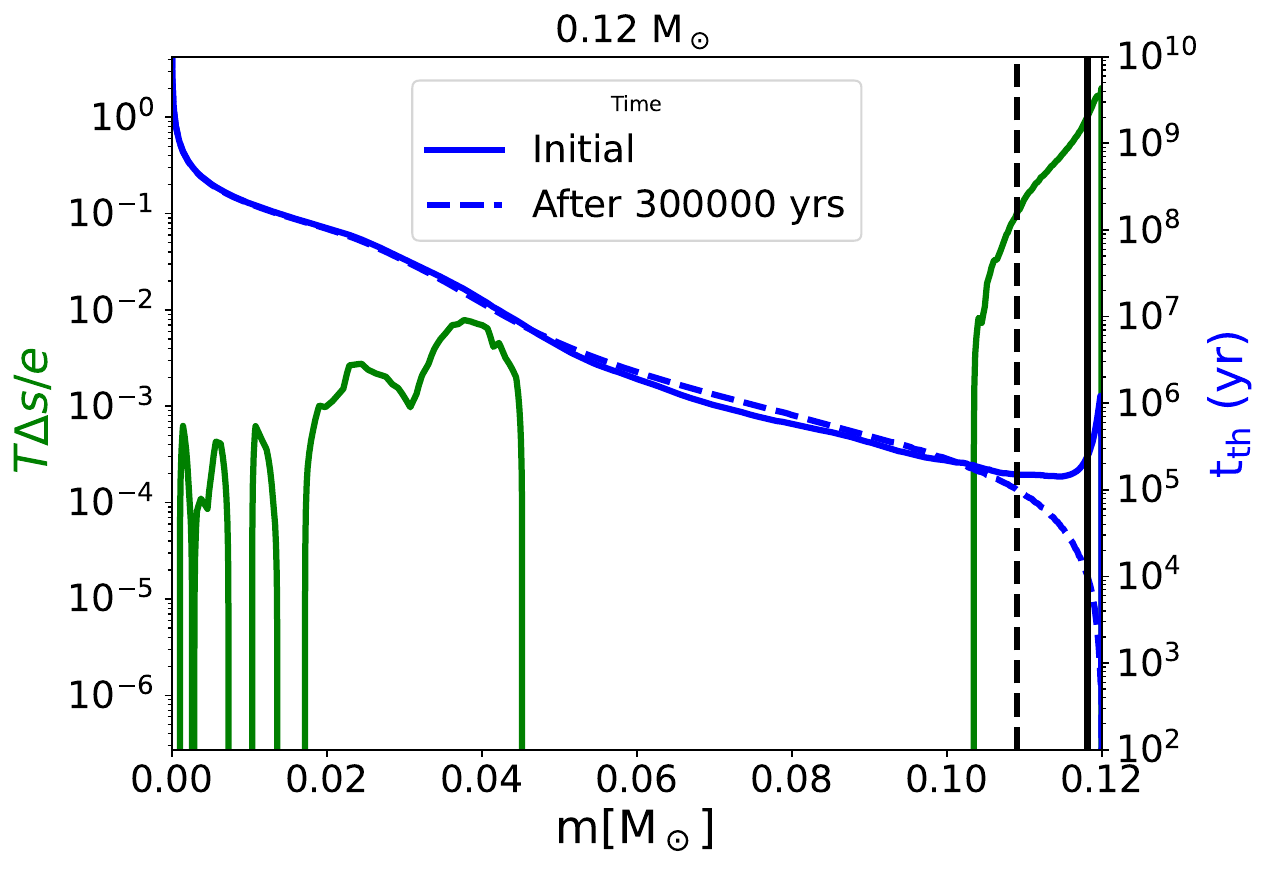}
\includegraphics[width=0.45\textwidth]{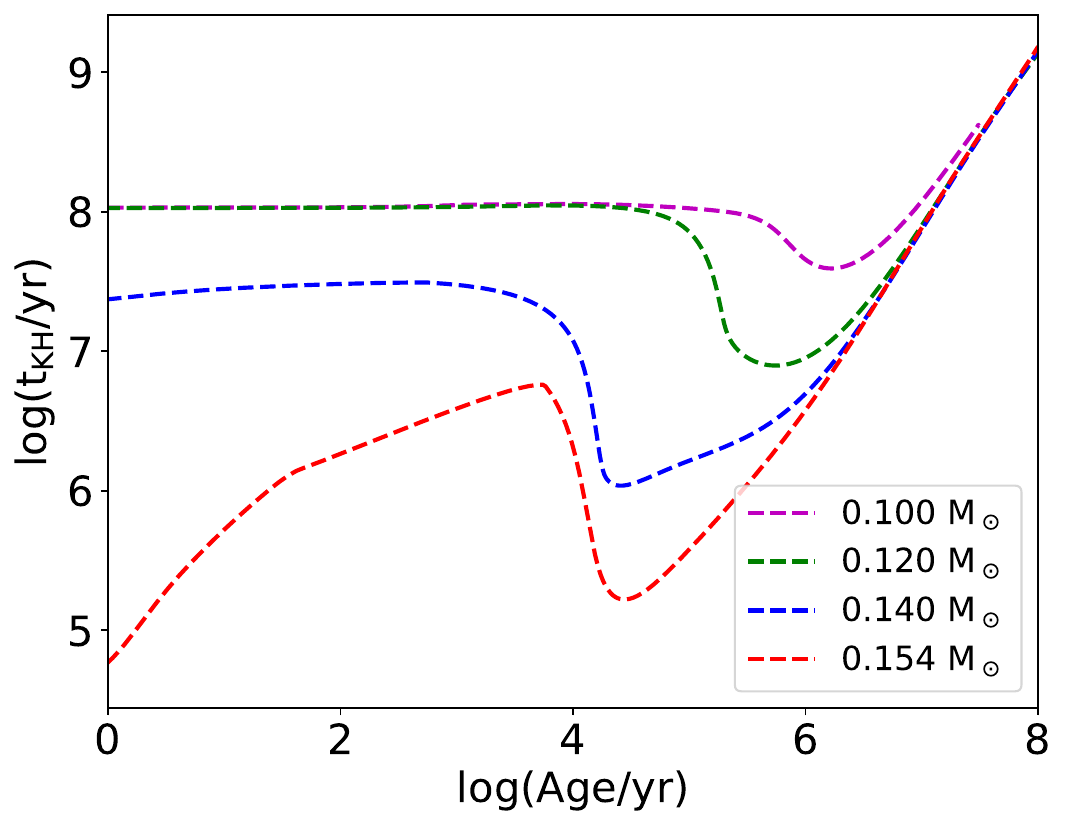}
\caption{[upper panel]: Local thermal timescale as a function of mass coordinate for the 4 models. [middle panel]: Heating fraction (green) and thermal timescale (blue) for the $0.12$ \msun\, model. [lower panel]: Kelvin-Helmholtz timescale of the models as function of age.}
\label{fig:th}
\end{figure}

The upper panel of Figure \ref{fig:th} shows the initial thermal timescales of the models. The sudden increase of the local thermal timescale in the outer layers is due to the surface contamination. The local pressure scale height and the opacity change the timescale associated with those layers. The middle panel shows an example of the 0.12 \msun\, model with an approximate estimate of local heating as a fraction of the internal energy $e$, $T\Delta S/e$ plotted alongside the thermal timescale. Unlike what was the case in \citet{Bhat1}, where the massive white dwarfs were only heated in layers with timescale of a few 1000 yr, the surface heating here is significant and the local thermal timescale is of the order corresponding to the kinematic ages of the observed runaways. The evolution of the Kelvin-Helmholtz time is shown in the lowest panel. This timescale is lower for the higher mass models which have a higher luminosity in the beginning. Once all the models have reached thermal equilibrium, this time is of the same order, at the beginning of the cooling track. This figure therefore shows that for these models the observed kinematic ages match the thermal timescales and can help explain why these stars are slightly inflated over the observed times.

\subsection{Surface Abundances}

\begin{figure}[!ht]
\centering
\hspace*{-0.2cm}
\includegraphics[width=0.45\textwidth]{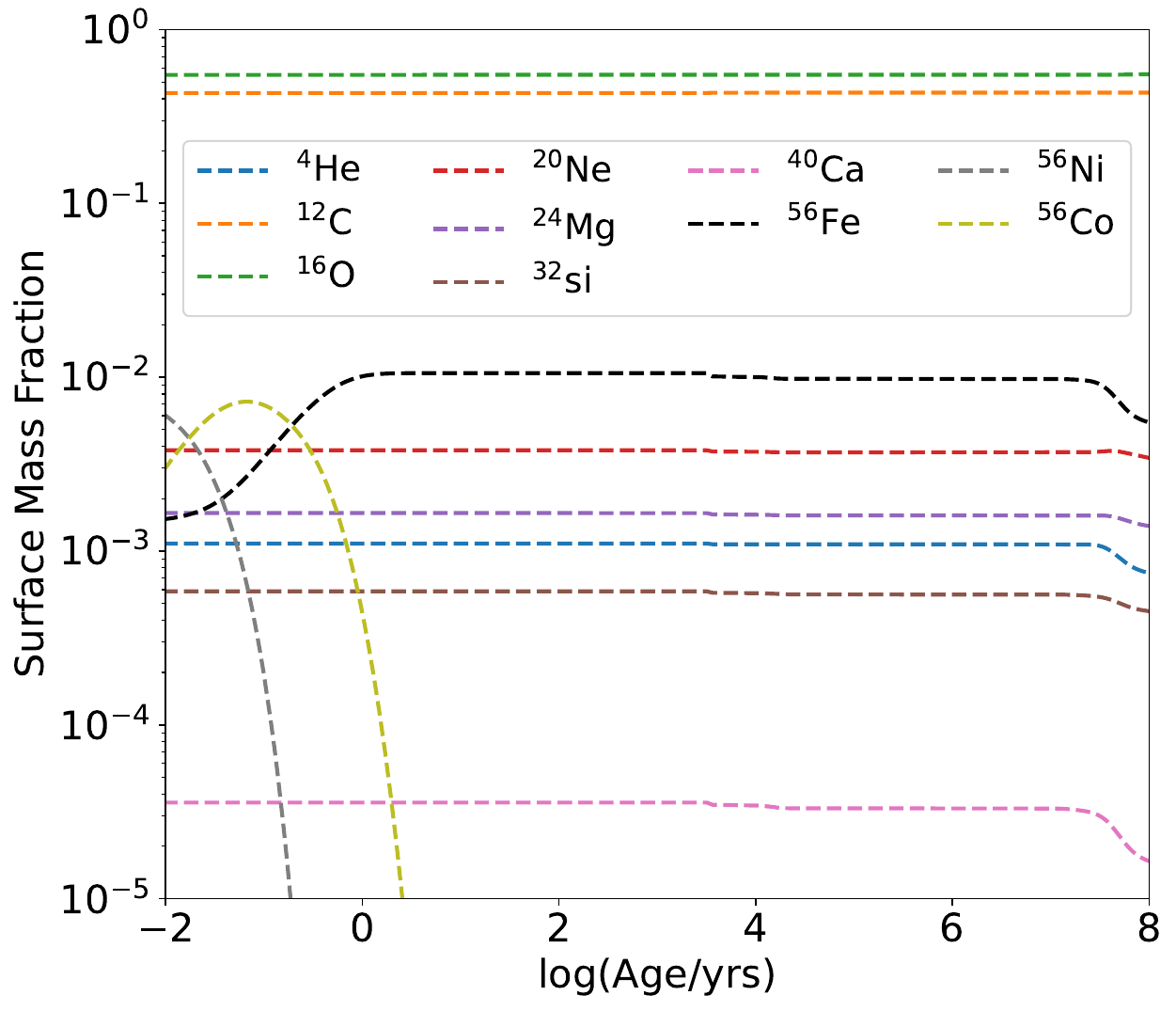} 
\caption{Mass fraction of elements which contribute significantly to the surface composition of the $0.154$ \msun model. The drop of these fractions in the later ages is due to thermohaline mixing.}
\label{fig:massfracs}
\end{figure}

Our lower mass models have allowed us to study in detail the heating and expansion of the runaway star due to the supernova shock and stripping. However, the surface abundances of such stars that have significant Ne, Mg, and Ca, apart from C and O, cannot be explained by those models. The 0.154 \msun\, model allows these elements to be studied directly. In Figure \ref{fig:massfracs} we show how the surface abundances evolve for our models.
%In this work, we have not utilized any element diffusion. For radiative levitation to dominate over gravitational settling, 
%\begin{equation}
%\frac{\kappa L}{4 \pi c R^2} \gtrsim \frac{G M}{R^2}
%\end{equation}
Within the temperature range of our models, iron opacities are of the order $0.01-0.1$ cm$^2$/g, and one needs a luminosity close to or higher than the Eddington luminosity for radiative levitation. This is different than the case of the hotter WDs from \citet{Bhat1} which were hotter than $10^5$ K, where the iron opacity bump decreases this luminosity to $10^2-10^3$ L$_\odot$. Therefore, for our models, thermohaline mixing and convection are the main sources of heat transfer and mixing throughout the star. %Since the envelope is expanded and non-degenerate the thermohaline mixing timescales are higher than what was observed in \citet{Bhat1}. 
%As such, we only used Type II opacities to calculate the evolution. %Nevertheless, hypervelocity white dwarfs provide an interesting avenue to study processes such as thermohaline mixing and radiative levitation and further modelling of these stars would benefit our understanding of these processes.
\subsection{Observations of Supernova remnants}

One of the key puzzles in the observations of hypervelocity stars from thermonuclear explosions is the lack of any runaways around known Type Ia supernova remnants \citep{2013ApJ...774...99K,2014ApJ...782...27K, 2018MNRAS.479..192K, 2022ApJ...933L..31S,shields2023}. One of the plausible solutions is that most Type Ia supernovae which occur in white dwarf binaries are quadruple detonations. In such a case the donor white dwarf also explodes due to the compression from the supernova shock igniting helium \citep{shen2024,boos2024}. However, if the violent merger scenario contributes $1-2\%$ to this population then a runaway must exist within $1-2\%$ SNR.

A second possibility is that a potential runaway was missed because all models so far predicted stars bigger than a solar radius and/or with surface temperatures greater than $10000$ K. Our model, on the contrary, shows that for the time when the runaway should still be inside the remnant ($\sim1000$ yr), it is actually under-luminous and cool compared to other models. They are less than a solar radius and $<10000$ K in surface temperature, leading to an average luminosity less than solar. In comparison, the models of \citet{Bhat1} and \citet{glanz1} are over-luminous by 4 orders of magnitude and hot enough to have significant UV emission. While optical and infrared observations might be helpful in the future, these observations could be contaminated by the supernova remnant itself.

\begin{figure}
\centering
\includegraphics[width=0.45\textwidth]{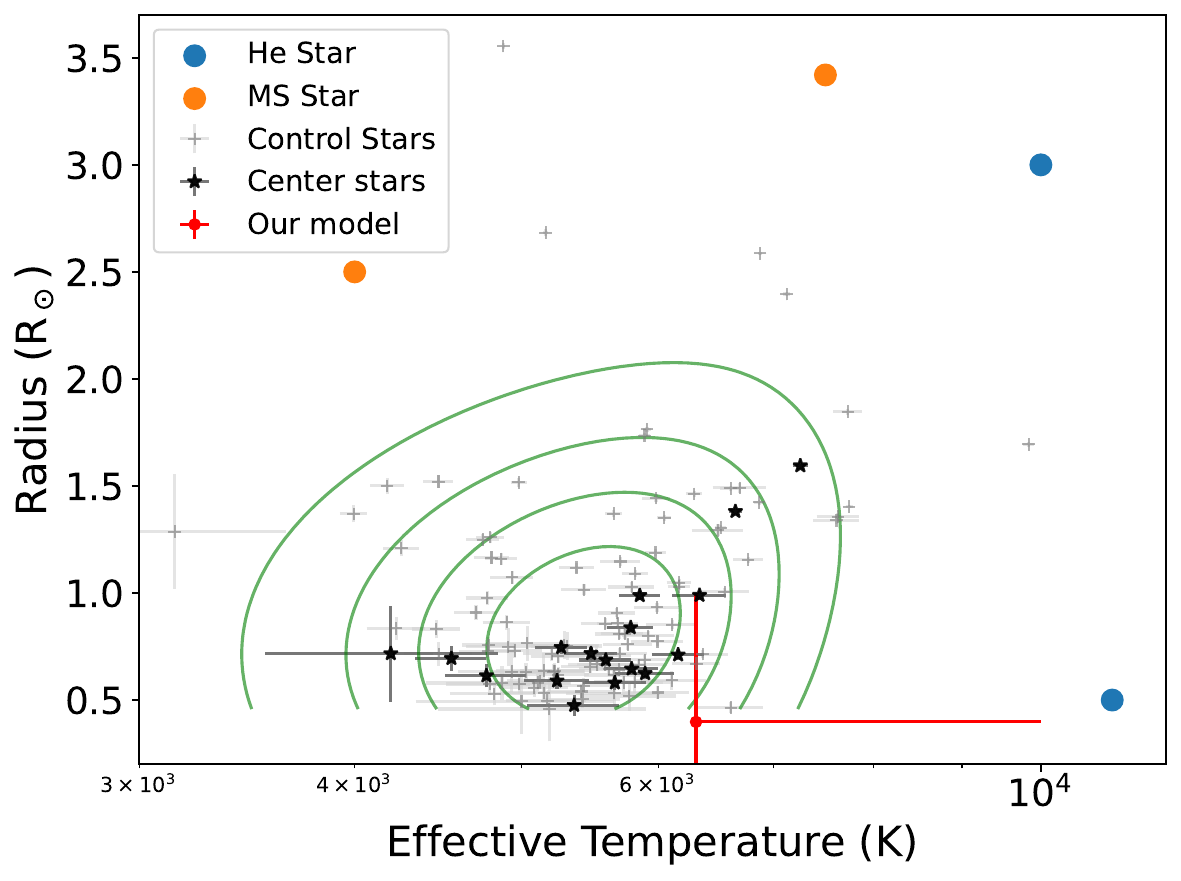}
\caption{The LMC SNR sample from \citet{shields2023}. Overlaid is the radius and effective temperature taking into account all four of our models in red, and the other He and MS star models which were discarded by \citet{shields2023}. }
\label{fig:snr}
\end{figure}

We recreate Figure 5 of \citet{shields2023} who compared the stars within supernova remnant SNR 0509-67.5 in the LMC with models of Type Ia-surviving donors. The control stars represent the main sequence outside of the central region while the centre stars were considered as possible runaways since they are close to the centre of the SNR. The models of \citet{Bhat1} and \citet{glanz1} that came after this study would also have been discarded as they were hotter and inflated. The recent models of \citet{wong2025} could be considered as long as the HeWD is less than 0.3 \msun. Our new model cannot be discarded right now as shown in Figure \ref{fig:snr}, and therefore provides a new avenue to study this (and similar) regions again.

\section{The formation of hypervelocity stars from white dwarf binaries}

\begin{figure*}
\centering
\includegraphics[width=0.95
\textwidth]{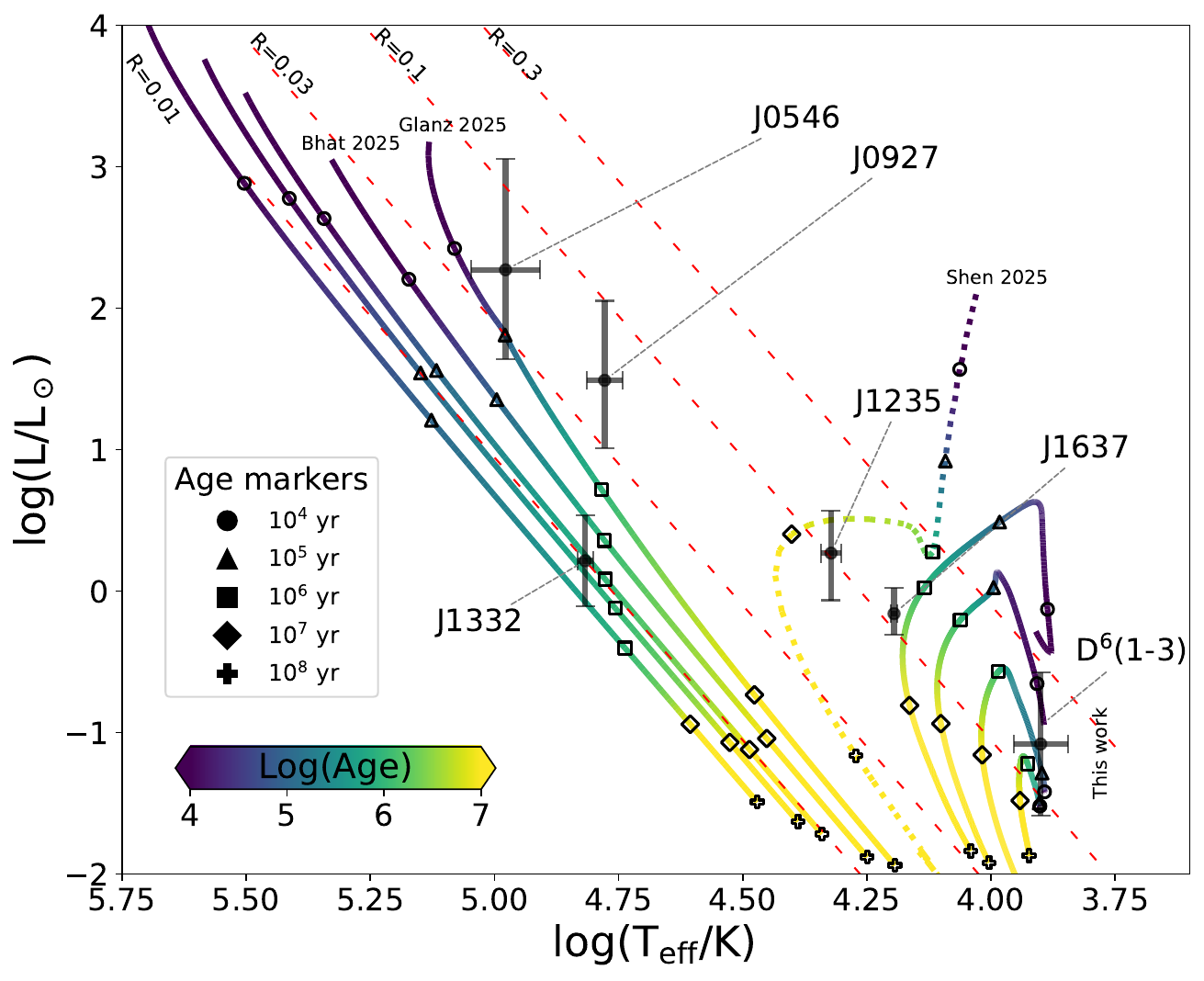}
\caption{HR diagram for all models. All observed D6 stars are plotted as black points. Dashed red lines represent lines of constant radii. Colorbar shows the log(age) of the star. The middle model going through J1235 is the 0.3 \msun\, model from \citet{ken2025}, plotted with a dotted line to differentiate from the other models for which a 3D simulation exists. }
\label{fig:HR_all}
\end{figure*}

We have shown in this work that the violent merger scenario is a possible way of producing cooler CO-dominated hypervelocity stars. For the hotter ones, merger-disruptions in HeCO WDs \citep{glanz1} and double detonations through Roche-lobe filling white dwarfs donors \citep{pakmor2022} have previously been studied. In Figure \ref{fig:HR_all} all of the known tracks from these three models have been plotted to create a unifying picture of the fate of the donor white dwarf in white dwarf binaries. To bridge the gap we have added one track of $0.3$ \msun\, taken from the recent work of \citet{ken2025}. Despite not originating from a 3D explosion simulation the model can be used to estimate the heating of heavier mass remnants from our merger scenario. For the purpose of clarity we only plot the tracks after the first 1000 yr of evolution, which are mainly dominated by super-Eddington expansion for the more massive white dwarfs. 

\begin{figure}
\centering
\includegraphics[width=0.45\textwidth]{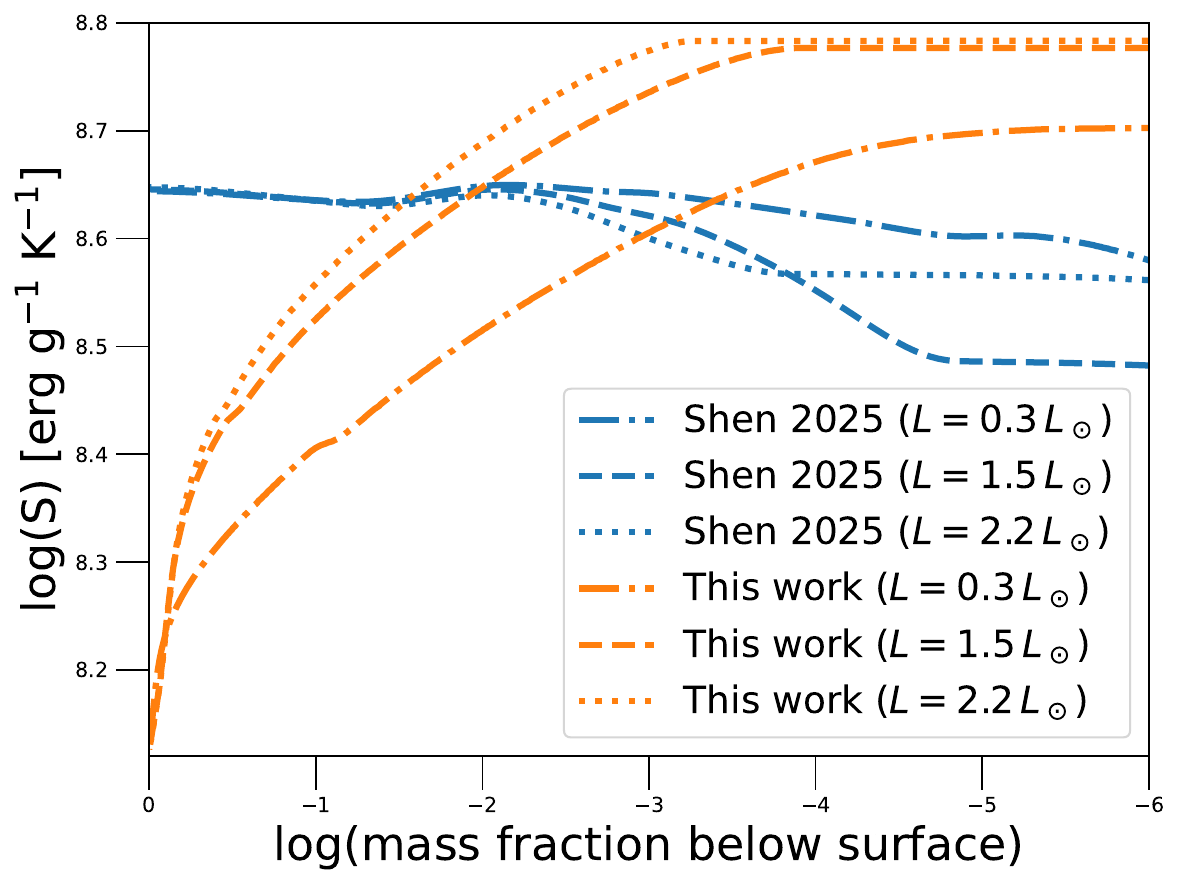}
\caption{Specific entropy as a function of the fractional mass below surface for our $0.154$ \msun\, model and the $0.154$ \msun\, fully convective model similar to the work of \citet{ken2025}. Three surface luminosities are plotted for comparison. }
\label{fig:comp_entropy}
\end{figure}

The main difference between our work and the models of \citet{ken2025} is that our 3D model shows that these stars are not completely convective and are partially degenerate. Figure \ref{fig:comp_entropy} compares the specific entropy of our $0.154$\,\msun model with a $0.154$ \msun\, fully convective star similar to those in \citet{ken2025} but calculated with MESA version 24.08.1 and the Type 2 opacities included with MESA, for a better comparison to the models presented in this work, at three different luminosities. Their models show a near constant inner entropy profile with changes occurring near the surface. Our models have a comparatively lower entropy near the center, but a higher entropy close to the surface. As such the local thermal time and not the Kelvin-Helmholtz time determine the inflated ages of the observed stars. 

In terms of the values presented here, assuming that J1332 is a hypervelocity star it may be explained by the double detonation scenario. No other hypervelocity runaway matches our models from the double-detonation scenario, unless the donors somehow lose more mass to be less massive than $0.5$\,\msun\, in the supernova that follows. This is highly unlikely, as the simulation of the $0.7$\,\msun\, white dwarf has shown previously. While the $0.5$\,\msun\, white dwarf has lower internal pressure it would be farther away from the accretor and therefore the shock would not necessarily strip away even more material. Beyond this limit, HeWD donors to massive CO WDs accretors may be able to explain the radius and temperature, but fail to explain the high speeds of the stars. HeWDs can only reach up to $\sim 1600$ \kms \citep{wong2025}.

The merger-disruption method of \citet{glanz1} offers a way to create the hotter runaways with increased speeds which stay expanded for long enough. Including our study, it is likely that these stars represent a lower mass end of the runaway distribution, whereas CO WD progenitors, as studied here could represent the higher mass end. Depending on the mass ratio and the amount of helium in the system mass loss could be significant to bridge the gap between 0.15 and 0.48 \msun. The new runaway J1637 as well as J1235 are both examples of systems which might have formed in such mergers. While not matching the kinematic ages exactly, the recent models of \citet{ken2025} show that is indeed possible.

For the case of merging white dwarfs, the Roche-lobe approximation for orbital velocities used by \citet{evan2021} and \citet{badry2023} underestimates their final velocities. This is because those calculations assume a supernova detonation when the donor WD has only just begun to overflow its Roche lobe. In contrast, merging WDs evolve significantly beyond this point, reaching closer to their companions and boosting their velocities further before detonation occurs. A better analytical velocity estimate can therefore be given by the circularization radius of the disrupted secondary. At Roche-lobe overflow the separation is given by $a_{\rm Roche}$. Under the assumption that the angular momentum at this point, dominated by the binary motion, is conserved, the circularization radius of the disrupted white dwarf is given by (see eq. 1 of \citet{fernandez2013})

%\begin{equation}
%\hspace{2cm}
%J_{\rm init} \;=\; \mu \sqrt{G\,M_{\rm tot}\,a_{\rm Roche}},
%\end{equation}
%\[
%\text{where }\; \mu=\frac{M_1 M_2}{M_{\rm tot}},\qquad M_{\rm tot}=M_1+M_2, \qquad q = \frac{M_2}{M_1}
%\]

%Under the assumption that the donor is fully disrupted and its mass forms a thin ring (torus) of radius $r_{\rm circ}$ around  $M_1$, we have, 
%\[
%\qquad
%\hspace{1cm}
%J_{\rm final} \;=\; M_2\,v_{\rm ring}\,r_{\rm circ}
%\;=\; M_2 \sqrt{G M_1 r_{\rm circ}}.
%\]

%Conservation of angular momentum implies $J_{\rm init} = J_{\rm final}$. This leads to $r_{\rm circ} \;=\; \frac{\mu^2\,M_{\rm tot}\,a_{\rm Roche}}{M_1\,M_2^2}.$

\begin{equation}
\hspace{1cm}
  r_{\rm circ} \;=\; \frac{M_1}{M_1+M_2}\; a_{\rm Roche} \;=\; \frac{a_{\rm Roche}}{1+q} 
\end{equation}

Based on this radius, we can calculate an upper bound of the orbital velocity for the donor at explosion as a function of the orbital velocity when the secondary first fills the Roche-lobe given in \citet{evan2021}.
\begin{equation}
v^{\mathrm{max}}_{\rm circ} = \sqrt{\frac{G M_1 }{r_{\rm circ}}} = (1 + q) \; v_{\rm Roche}
\end{equation}

However, a major assumption in the derivation above is that the secondary is completely disrupted to form a torus around the primary. This is obviously not the case in reality, since an explosion happens before such a state can be reached. Therefore, the center of mass (COM) of the system is still not at the center of the primary but close to it. A simple analytical expression to calculate the ejection velocity as the orbital velocity with respect to the COM is therefore given by assuming that the distance of the secondary to the COM is approximately $r_{\text{circ}}/(1+q)$, leading to,

\begin{equation}
v_{\rm circ} = \sqrt{\frac{G M_1 (1+q) }{r_{\rm circ}}} = \sqrt{(1 + q)} \; v_{\rm Roche}
\end{equation}

This is equivalent to
\begin{equation}
v_{\rm circ} = 
\sqrt{
\frac{
0.49\, q^{2/3}\, \, G M_1
}{
R_2 \left[ 0.6\, q^{2/3} + \ln(1 + q^{1/3}) \right]
}
}
\end{equation}

\begin{figure}
\centering
\includegraphics[width=0.45\textwidth]{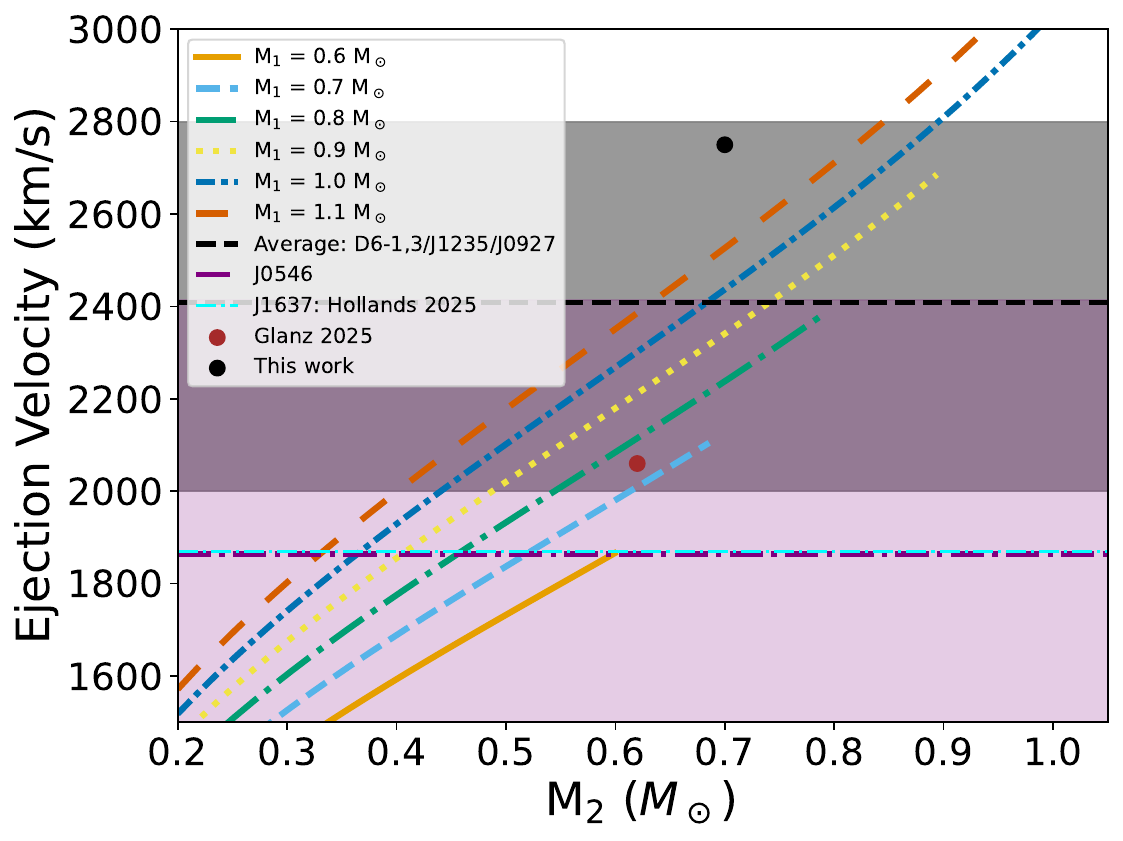}
\caption{Orbital velocities for multiple primary and secondary masses (the lines correspond to different primaries). The velocity is calculated assuming circularization radius is reached before explosion. Two simulation values are plotted as circles. Black region shows min and max ejection velocities for D6-1,3, J1235, and J0927. Cyan lines show the values for J1637 and J0546, both matching slightly lower mass white dwarf secondaries. }
\label{fig:vorb}
\end{figure}

This velocity is plotted in Figure \ref{fig:vorb} in a similar manner to \citet{glanz1}, although the assumption was that there was tidal disruption and the mass range was relatively smaller. The figure shows clearly that the orbital velocities allowed for merging white dwarfs can be much higher than previously assumed. The slight excess velocity in the simulation is most likely due to the final surviving part of the white dwarf being even closer than this radius. Furthermore, any angular momentum loss will result in an even closer object, which is neglected in the analytical calculation. Since the calculation only works for the COM, part of the donor would plunge in faster than this velocity. 

The calculation to explain the significant mass loss is not as straightforward but can be motivated by previous studies \citep{evan2019,wong2025}. The ejecta from the supernova hit the donor ($M_2$) with a ram pressure equal to

\begin{equation}
\hspace{2 cm}
    P_{\mathrm{ej}}\approx \frac{M_1 v^2_{\mathrm{ej}}/2}{4\pi a^3/3}
\end{equation}

Since $a$, close to disruption, scales inversely with donor mass, massive donors with massive accretors will experience higher ejecta ram pressure. At the separation we have calculated above, the ejecta pressure would be $(1+q)^3$ times more than the pressure when they start to fill their Roche lobes (as was the case for the D6 scenario in \citet{pakmor2022}). A second important quantity is the pressure within the donor, which may be characterized by the central pressure, which scales with $M^2_2/R^4_2$. A consequence of this is that despite feeling a higher pressure, the corresponding shock strength is lower in more massive CO white dwarfs, compared to hot subdwarfs and He white dwarfs \citep{evan2019,wong2025}. This leads to a lower fraction of mass loss and shock heating in more massive white dwarfs \citep{Bhat1}. 

However, white dwarfs close to disruption have significant rotational energy, are tidally heated, and significantly distorted. Due to the dynamically unstable nature of mass transfer in white dwarfs the secondary will also be puffier due to mass loss before the explosion. This can change the central pressure and deviate significantly from the simple polytropic consideration. The inverse dependence of the pressure on $R^4$ also causes this pressure to drop very quickly. This is also clear from Figure \ref{fig:rho_T}, where even after a few $1000$ s the runaway star is only partially degenerate. Furthermore, from our simulation the center of the secondary goes from $> 10^8$ g/cm$^3$ at the beginning of the simulation to $\sim10^7$ g/cm$^3$ before the explosion. A similar drop in central pressure is seen such that the ejecta pressure is of the same order as the central pressure. If the outer layers of the white dwarfs are close to complete disruption, then the mass loss will be set by the ram pressure due to the incoming ejecta, the coupling of this ejecta with the secondary, and how disrupted the secondary white dwarf is at that moment. Therefore, as the donor and accretor masses increase, the donor could in some cases lose significantly more of its mass, and therefore lower mass runaways could be expected to be formed. This is probably why our system ends up with a lower mass runaway than the one in \citet{glanz1}. Future simulations should help shed more clarity on this question.

\section{Conclusions}

In this paper we have presented results for the first time of a 3D simulation and corresponding 1D stellar evolution of a runaway star produced from the violent merger supernova scenario. In the cases when the primary white dwarf does not undergo double detonation to create a supernova, the secondary white dwarf can continue mass transfer and come closer to the primary. On its last orbit, close to the circularization radius of the system, the secondary is almost completely distorted and the resulting infall ignites carbon on the primary white dwarf and destroys it. In the aftermath, a relatively low mass, 0.16 \msun\, chunk of the donor flies away at a speed roughly ($\sqrt{1+q}$) times the orbital velocity at the beginning of Roche-lobe overflow. 

We mapped the simulation results into MESA to evolve heated white dwarfs representing surviving chunks of pre-supernova donors in the mass range $0.10\,$M$_\odot - 0.156\,$M$_\odot$. We then evolved these models to 100 Myr after the supernova explosion. Our results match the observed values of the cooler stars quite well. In particular, the thermal times of these models match the kinematic ages of the observed hypervelocity stars. With future studies, the parameter range of the violent merger scenario and the resulting runaway masses must be explored.

Our study reopens the possibility to revisit observations of supernova remnants for hypervelocity elopers. Since these stars are not as hot and contaminated as previously believed, they might be hiding in the surrounding main sequence population.
\section*{Acknowledgements}

This project succeeded a project which was originally started as part of the Kavli Summer Program which took place in the Max Planck Institute for Astrophysics in Garching in July 2023, supported by the Kavli Foundation. AB is extremely grateful to Stephen Justham and Selma de Mink for not listening to him back then, and assigning him the runaway project. A.B. was supported by the Deutsche Forschungsgemeinschaft (DFG) through grant GE2506/18-1.
  K.J.S.\ was supported by NASA through the Astrophysics Theory Program (80NSSC20K0544) and by NASA/ESA Hubble Space Telescope programs \#15871 and \#15918.
Work by EB was performed under the auspices of the U.S. Department of Energy by Lawrence Livermore National Laboratory under Contract DE-AC52-07NA27344.

\clearpage
\appendix



\clearpage
\bibliographystyle{aa}
\bibliography{main} % if your bibtex file is called example.bib

%\begin{thebibliography}{99}
%\end{thebibliography}

%%%%%%%%%%%%%%%%%%%%%%%%%%%%%%%%%%%%%%%%%%%%%%%%%%

%%%%%%%%%%%%%%%%% APPENDICES %%%%%%%%%%%%%%%%%%%%%

%\appendix

%%%%%%%%%%%%%%%%%%%%%%%%%%%%%%%%%%%%%%%%%%%%%%%%%%

% Don't change these lines (for MNRAS)
%\bsp	% typesetting comment
%\label{lastpage}

\end{document}